\documentclass[aps,prl,twocolumn,showpacs,superscriptaddress]{revtex4}
  \usepackage{color,graphicx,shortvrb,latexsym}%

\begin{document}

\title{\bf Giant microwave photoresistance of two-dimensional electron 
gas }

\author{ P. D. Ye} 
\affiliation{ National High Magnetic Field Laboratory, Tallahassee,
Florida 32310} 
\affiliation{ Department of Electrical Engineering, Princeton University,
Princeton, New Jersey 08544} 
\author{L. W. Engel}
\affiliation{ National High Magnetic Field Laboratory, Tallahassee,
Florida 32310}
\author{D. C. Tsui}
\affiliation{ Department of Electrical Engineering, Princeton University,
Princeton, New Jersey 08544}
\author{J. A. Simmons}\author{J. R. Wendt} \author{G. A. Vawter}\author{J. L. Reno} 
\affiliation{ Sandia National Laboratories, Albuquerque, New Mexico 87185}

 \date{\today}

\begin{abstract}
 We measure  microwave frequency (4-40 GHz) photoresistance at low  magnetic 
 field  $B$,  in  
 high mobility 2D electron gas samples,  excited by signals applied 
 to a transmission line fabricated on the sample surface. Oscillatory   
 photoresistance vs $B$  is observed. For excitation at the cyclotron 
 resonance  frequency, we find an unprecedented,  giant relative photoresistance $\Delta R/R$ of 
 up to 250 percent.   The photoresistance is apparently proportional to  the square root of 
 applied power, and disappears  as the temperature is increased.

\end{abstract}
 
\maketitle

The two dimensional electron gas (2DEG) in GaAs/Al$_x$Ga$_{1-x}$As 
heterojunctions has 
long been known to exhibit resonant photoresponse to 
  illumination at the cyclotron resonance frequency, $f_C$.   This 
  resonant  response  has generally 
been attributed to  heating of the electrons by absorption of  energy 
from the radiation field.     
The earliest photoresistance experiments on 2DEG, in both the far 
infrared (FIR)\cite{maantsui,horstman} and millimeter wave\cite{vasilnodots} 
regimes,  showed    single, well-defined positive photoresistance 
features   near the  cyclotron resonance condition,  $f =eB/2\pi m^*$,
where $f$ is the applied frequency and $m^*$ is the effective mass in 
GaAs, and $eB/2\pi m^*$ is the cyclotron frequency, $f_C$. 
$\Delta R/R$, the change of  dc resistance due to applying  radiation,
 divided by the resistance with the radiation off, was much less than 
 unity.

 Later   work\cite{zudov}  for $f=30$ to 150 GHz,     used higher mobility 
  samples ($\mu\sim 3\times 10^6\rm 
  cm^2/Vs$, about three times that in the earlier 
  investigation\cite{vasilnodots}), and  revealed  
    $\Delta R$  of  large  relative size, with  larger $\Delta R/R\approx 0.35$.  
Strikingly, $\Delta R$ vs $B$ showed   oscillations with alternating positive and 
negative $\Delta R$.  These $\Delta R$ oscillations  were periodic in $1/B$ like  
Shubnikov-de Haas (SdH) oscillations, but  had  $f$-dependent 
 period $e/m^*2\pi f$. $\Delta R$ peaks occurred at $f\approx j f_C$, for 
$j=1,2,3,,,$.   The $j=1$ peak was largest   but peaks  for $j$ up to 7 were 
observed.    To explain the   oscillatory 
photoresistance  the authors of ref. \cite{zudov}, 
proposed a process  in which 
impurity scattering  combines with  microwave absorption at $j f_C$.  
This causes  transitions between Landau orbitals 
with  energy   quantum numbers 
differing by $j$ and  guiding centers displaced from each 
other.
 
In this paper we report on    
photoresistance  measurements,  in which microwaves were  applied to a transmission line 
made of metal film  on the top surface of the sample.   We find   
an oscillatory $\Delta R/R$  whose maximum  value is 2.5, nearly 
an order of magnitude larger than previously observed\cite{zudov}. 
The giant photoresistance   decreases with $T$, and is 
proportional to the incident microwave {\em amplitude}, that is, to the 
square root of the applied power.

We performed measurements on samples originally designed for 
broadband, transmission line based  
measurements of the microwave conductivity of the 2DEG\cite{engel}.
Fig.~1  is a schematic illustration of a sample.    On the top  surface 
of the 3 $\times$ 5 mm sample, a film of 
200  \AA\   Ti and 3500  \AA\    Au was patterned  to 
form a coplanar waveguide (CPW)\cite{wen} transmission 
line, with  a 30 $\mu$m wide center strip separated from 
side ground planes by slots of 20 $\mu$m-width, which are  shown in light grey 
in the figure.     
In the limit of low 2DEG conductivity, the in-plane microwave fields
produced by driving the CPW are  well-confined within the slots; in 
the present experiments some of the microwave field leaks 
under the metal film near the slot.    We 
studied the change, on applying microwaves to the CPW, of  
quasi-dc resistance measured using combinations of the  alloyed AuGe/Ni 
 contacts at the edges, marked 1 through 8 in the figure. 
The 2DEG under the tapers of the CPW at the edges of the 
sample (the areas indicated by the dashed lines in Fig.~1) was removed 
by chemical wet etching.  The rest of the   sample was not etched,
so the ohmic contacts are connected to the slot areas  by the 2DEG under the 
ground planes. 

We measured two samples, both made from the same wafer,
a high-mobility GaAs-AlGaAs 
heterojunction  in which the 2DEG was located approximately 120 nm 
underneath the sample surface.  
Sample 2, but not sample 1, contained an antidot 
array, which was placed  just in the  two  CPW slots.  
To produce the antidots, 500 nm period square lattices of 
 50 nm diameter, 50 nm deep holes were defined  by electron beam lithography  
and   reactive ion beam etching\cite{vawter}. 
The samples were prepared for measurement with a brief illumination 
from a red LED, after which 0.3 K mobility (without antidots)
was around  3 to $4 \times 10^6 \, \rm cm^2/Vs $.  As determined from 
Shubnikov-de Haas (SdH) oscillations in $R$ vs $B$, the densities of 
sample 1  and sample 2 were respectively 1.7 and 2.1 $ \times 
10^{11}$cm $^{-2}$.

The quasi-dc  resistance measurements  used 
 typical operating frequency of 13.5 Hz, and applied current of 100 nA rms.
We report resistances, $R$, measured  from four contacts along one 
edge of the sample, in the  topology associated with diagonal 
resistivity, with 
injected current between contacts 1 and 4 on Fig.~1 , and  voltage taken 
between contacts 2 and 3. 
 With the microwaves  both off and on, some asymmetry of $R$ between positive and 
negative $B$ was present in sample 1 , and to a much lesser extent in sample 2.  
We attribute this asymmetry to 
inhomogeneity produced by illumination, and to the large size and close 
proximity  to each other of the contacts. All the  $R(B)$  presented 
here   are  the average of  data taken at  
$B$ and $-B$.

The samples were measured in  a dilution refrigerator cryostat. 
A Hewlett Packard 8722D network 
analyzer at room-temperature  supplied microwave signals to  cryogenically 
compatible coaxial cables that were connected to the CPW.
The external magnetic field ($B$) 
was   normal to the plane of the 2DEG.

Fig.~2a shows $R$  vs $B$  for sample 1. Along with a reference trace 
taken with the microwave source off, traces are shown for 
 several applied microwave frequencies, $f$.
The  microwave power $P$ was 100 nW  at 50 ohms, incident onto the edge of the CPW. 
The ``bath'' temperature (that of the metal on which the sample was 
mounted)   was about 100 mK,  but heating due to the applied 
power must have 
occurred.   
When microwaves are incident on the CPW, $R$ vs $B$ exhibits a series 
of peaks.  As $f$ increases, 
the oscillations shift  to higher $B$ and grow 
 stronger,  and more  features  emerge at lower $B$.  For $f\ge 20 $ GHz 
there are three peaks,  which we will see have roughly even spacing in  $1/B$.

Data for sample 2 (with antidots in the CPW slots) 
 are shown in Fig.~2b.   The $1/B$ oscillations exist similar to sample 1.
  The different  behavior  of the two samples,  both with and   without the 
microwaves,   are likely due to  different red-light illumination 
doses rather than to the presence of the antidots in sample 2. 
The antidots   have little effect on the measurement 
  since the measuring contacts are all on one side 
   of  the antidot strips and  so are somewhat remote from them. 

Our main result is the {\em large size} of 
the photoresistance seen particularly on the highest-$B$ peak. There  
we find $\Delta R/R$, the change 
in $R$ on applying microwaves divided by the microwaves-off value, as 
high as  2.5 for sample 1 and 2.0 for sample 2.
The large   $\Delta R/R $ cannot be explained by $R $ being made small 
by  cancellation of Hall  and diagonal contributions to the 
measured resistance, so it is clear that the application of the  
microwaves near resonance produces a drastic  change in the transport 
of the 2DEG.

 The  photoresistance oscillations   are 
 related to $f$ as described in ref. \cite{zudov}; with $f=jf_C$, maxima occur at 
  integer $j$, and minima at half-integer $j$. 
Fig.~3a shows the   positions of the $j=1$ and $j=2$
microwave-produced  maxima of  both samples.   The applied frequency, $f$, is plotted against the $B$ at which the 
maximum occurs.    $f_C$ and $2f_C$ are also 
plotted, where we have used $m^*$ of 0.07  times the free electron mass. 
The maxima falling close to the lines demonstrates the approximate 
$1/B$ peridicity of the photoresistance oscillations.
The highest $B$ ($j=1$) maximum
 falls at significantly higher 
$B$ than the $f_C$ line, especially for sample 2. 
The    $j=2$ maximum  falls on  $2f_C$ to 
within about 5 percent for both samples.  
Outward  shift of the first harmonic was also reported  in ref. \cite{zudov}.
 
  In Fig~3b, 
we plot  $\Delta R$ vs $T$ for  the  $j=$1,3/2, and 2
photoresistance extrema of  sample 2, for $P\approx 100$ nW, at $f=30$ GHz. $\Delta R$ is negative  for 
$j=3/2$, and positive for $j=1$ and 2.   
$|\Delta R|$ increases with decreasing $T$,  saturating around   500 
mK.  30 GHz photons have energy   $k_B T$ at 1.43 K;   
 $\Delta R$ vs $T$   is significantly reduced from its maximum by that 
 temperature, and is likely   characterized by  that energy.
 
 Fig.~3c shows $\Delta R$ vs $P$, the     microwave power incident onto 
the CPW, for sample 2 with $f=30$ GHz, at the $j=1$ and $j=2$ maxima 
 and the $j=3/2$ minimum.       The ``bath'' temperature, $T$, to which the 
electrons  cool in the limit of very small $P$ applied, was around 
100 mK. Lines with  
  $\Delta R\propto P^{1/2}$ appear on the graph; this behavior fits the 
  $j=1$ data well and is at least consistent with the data for
  $j=3/2$ and $j=2$.   The $\Delta R\propto P^{1/2}$  behavior  appears to hold well down to small 
  $P$, where heating of the sample  would be much less.  
$\Delta R$ vs $P$ may be affected  
 by    (1) heating of the sample
  and  by (2) reflectance of   the CPW varying with $P$.
  The second effect 
 could make the microwave intensity  that reaches the relevant area of 2DEG  
 nonlinear in   the power incident onto the CPW.

The samples we looked at have similar mobility to those examined in ref. 
\cite{zudov}. The high ends of our  $f$ and $T$ ranges overlap the conditions 
studied   in that reference, for example at $f=30$ GHz, $T=2$ K.   The  much larger 
$\Delta R/R$ presently observed may be due to a larger power 
density reaching the 2DEG, or to inhomogeneity caused   by the opaque  
ground planes upon illuminating the sample with red light.
 
In conclusion, we have found  a low $B$   microwave photoresistance
  much larger than previously observed; the features  
 have apparent  $P^{1/2}$ dependence, and  a characteristic 
$T\sim hf/k_B$.

We would like to thank W. E. Baca,  
J. Wang and J. Yao   for  technical assistance and C. M. Hu for valuable discussions. We 
especially thank  R. R. Du for  making ref. \cite{zudov} available before publication. 
This work is supported by the 
Air Force Office of Scientific Research, and the National Science 
Foundation.

\newpage\ \\ \newpage\ \\

\begin{figure}[ht] 
 \noindent\includegraphics[scale=3]{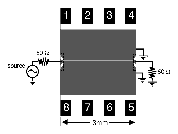} 
\caption{ \ \ \   Schematic view of sample and microwave connections.
Dark gray represents metal film, numbered black  rectangles represent 
ohmic contacts.    The slots are shown in light gray, 
and contain antidots for sample  2 only.  The sample is 5mm long by 
3mm  wide,  and the edge of the ground plane  is  about 1 mm from the top edge of 
the sample.}   
\end{figure}

 \begin{figure}[ht] 
 \noindent\includegraphics[scale=4]{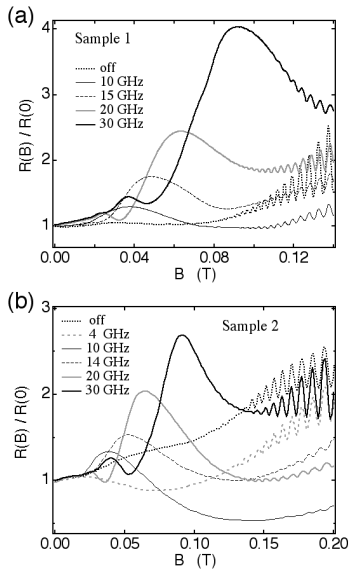}\\
\caption{   \ \ \ Resistance $R$ vs magnetic field $B$, 
for  microwave excitation  at various frequencies $f$. Data are 
normalized to $B=0$ value $R(0)$.   Reference  
traces taken with microwave power off are also included.} 
\end{figure}  
 \newpage\ \\

\begin{figure}[ht]  
 \noindent\includegraphics[scale=1.1]{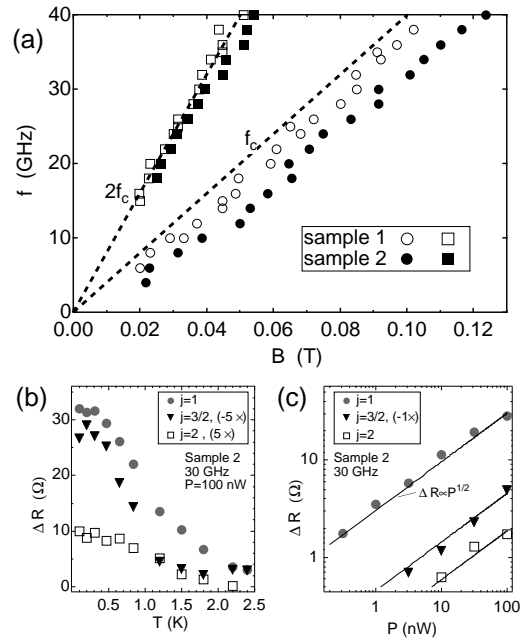} 
\caption{  {\em a.}\ \ \  Microwave excitation frequency ($f$) plotted 
vs  magnetic field ($B$) position of two largest photoresistance 
maxima.\ \ {\em b.}\ \ \  30 GHz 
$\Delta R$, for $P= 100$ nW, vs cryostat temperature $T$.  
$\Delta R $  is taken at  maxima with $j=1,2$, and the 
minimum  at $j=3/2$, where $f=jf_C$, and $f_C$ is the cyclotron frequency.  {\em c.}\ \ \  Photoresistance $\Delta R$ for microwave 
frequency $f= 30$ GHz, vs power ($P$) at 50 $\Omega$ incident on edge of 
sample.   Cryostat temperature was $T\sim 100$ mK.  Lines show best 
fits to $\Delta R  \propto P^{1/2}$.\ \ \  }
\end{figure}   
  
\end{document}